\documentstyle[11pt,twoside,sympo2011,epsfig]{article}
\begin{document}
\title{Seismological Diagnostics for Solar-like Stars}
\author{Ian. W Roxburgh$^1$}
\affil{$^1$ Queen Mary University of London  [i.w.roxburgh@qmul.ac.uk]} 
\begin{abstract}
The oscillations in solar like stars are described in terms of  the phase shifts of the eigenmodes from simple sine-waves. We discuss model fitting and inversion techniques based on this representation. We analyse the periodic signatures from the HeII ionisation zone and base of the convective envelope of the CoRoT star  HD49933.

\end{abstract}

\section{Phase shifts and diagnostics}
The oscillation frequencies $\nu_{n,\ell}$ of a spherical star can always be written in the form
\vskip-3pt
$$\nu_{n,\ell} =\Delta \big[n+{\ell/ 2} + \beta_{\ell}(\nu) \big]\eqno(1)$$
\noindent where $\Delta$ is a constant and $\beta_\ell$ are functions of $\nu$.
For $\beta_\ell=0$ the frequencies are uniformly separated, which is the asymptotic behaviour of the frequencies of a uniform isothermal sphere. 
For a real star the residuals $\beta_{\ell}(\nu)$ to a simple fit to $\Delta(n+\ell/2)$ contain all the information on the departure of the frequencies from uniform spacing, and hence on the departure of the structure of the star from that of a uniform sphere. The $\beta_\ell$ are determined by integrals of the structure variables over the star;  these integrals can be split into a contribution $\alpha_\ell$ from outer layers and $\delta_\ell$ from the inner layers and Eqn 1 expressed as (Roxburgh \& Vorontsov 2000, Roxburgh 2009)
\vskip-3pt 
$$\nu_{n,\ell} =\Delta_T \left(n+{\ell\over 2}\right) + {\Delta_T\over\pi} \big[\alpha_{\ell}(\nu)-\delta_{\ell}(\nu)\big],~~~
{\rm where~~~}\Delta_T={1\over 2T}~~~{\rm and}~~~~T=\int_0^{R_a} {dr\over c}\eqno(2)$$
\noindent is the acoustic radius of the star ($c$ the sound speed). The phase shifts $\alpha_\ell, \delta_\ell$ for a solar model are shown in Fig 1. Note that $\alpha_\ell$ is essentially independent of $\ell$, and that both $\alpha(\nu,t)$ and $\delta_\ell(\nu,t)$ are almost constant in the intermediate layers of the star. 

With $\alpha_\ell(\nu)$ independent of $\ell$, the differences $\beta_\ell-\beta_0$  give $\delta_\ell-\delta_0$, a function of $\nu$ determined solely of the interior structure. One way to do this is to take the ratio of small to large separations (R\&V 2003a) which can then be used to find a best fit model to an observed frequency set. The model filtering algorithm 
(R\&V 2003b, R 2010) is based on the  deduction from Eqn 2 that $\delta_\ell(\nu, t) -\pi\Delta(n+\ell/2)+\nu\Delta/\pi$ must collapse to a function only of $\nu$ in the outer layers of a star; by calculating this function for each model, using an observed frequency set, we can determine which model best satisfies this condition (see Fig 2). This also gives an inversion algorithm in which the model is iteratively corrected to improve the fit (Vorontsov 2001, R\&V 2002, 2003b).

The dominant variation of $\beta_\ell(\nu)$ is due to the internal structure but sharp structural changes produce a quasi periodic modulation in $\beta_\ell(\nu)$ with periods of twice the acoustic depth and acoustic radius of the layer, the 2 frequencies satisfying $1/\nu_1 + 1/\nu_2=2 T$. On subtracting off the smooth variation in $\beta_\ell(\nu)$ by a best fit low order polynomial we can fit the residuals to harmonic functions to obtain the modulation frequencies and hence acoustic depths.
Figure 3 shows the result of this procedure applied to the  CoRoT frequency set for HD49933  (Benomar et al (2009). The plots show a local maxima at modulation frequencies of $\sim620$ and $\sim360\mu$Hz, corresponding to acoustic depths of $\sim800$ for the  HeII ionisation layer  and $\sim1400$ secs for the base of the convective envelope, which should be 
compared with the estimated acoustic radius of $5800$ secs ($=1/(2\Delta$). This provides significant constraints on models of HD49933.

\newpage

\begin{figure}[h]
\vskip-20pt
\begin{center}
   \epsfig{file=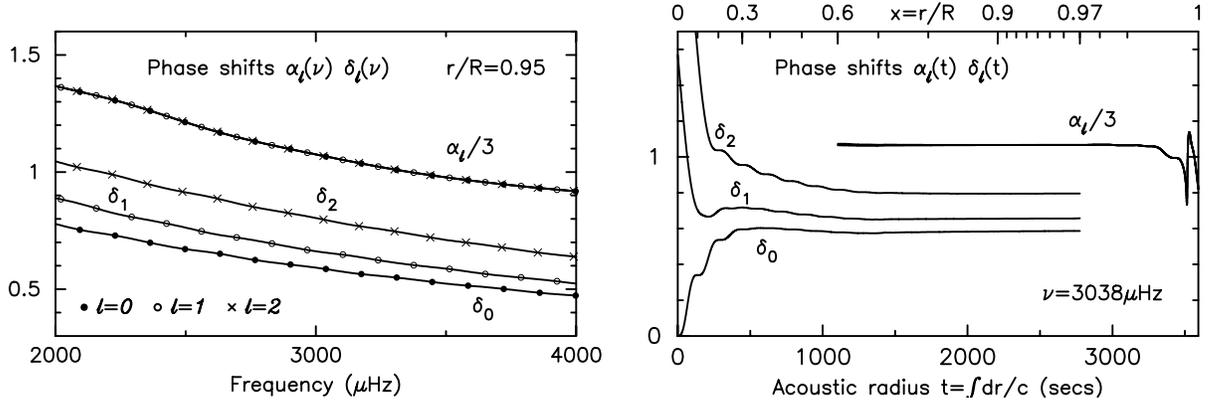, width=5.3cm,angle=270}
   \vskip-5pt 
\caption{Phase shifts $\alpha_\ell, \delta_\ell$ as a function of frequency;  and their variation  within the star. }
\end{center}
\end{figure}
\vskip -25pt
\begin{figure}[h]
\begin{center}
    \epsfig{file=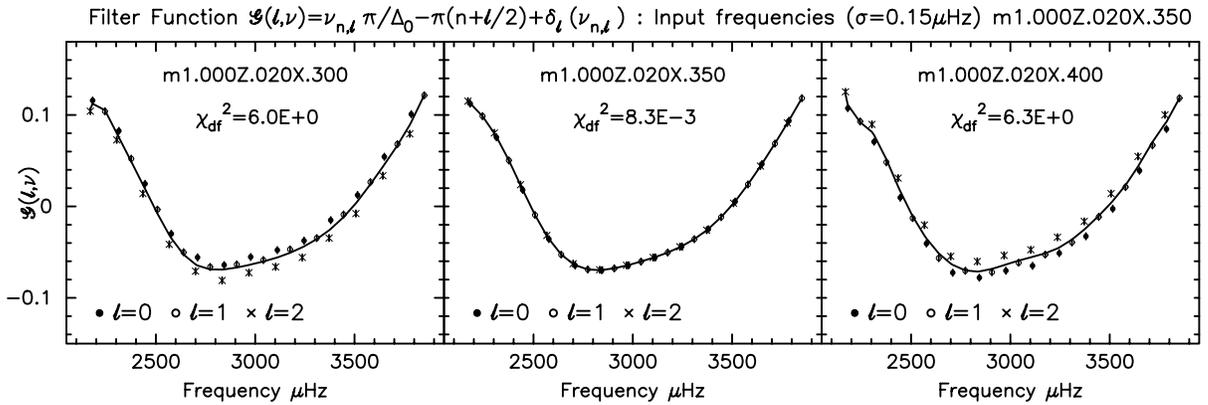,width=5.3cm,angle=270}  
   \vskip-5pt 
\caption{Filter function ${\cal{ G}} (\ell,\nu)$ and goodness of fit to a function of $\nu$ ($\chi^2$ per degree of freedom)}
\end{center}
\end{figure}
\vskip-15pt
 \begin{figure}[here!]
   \centering
   \epsfig{file=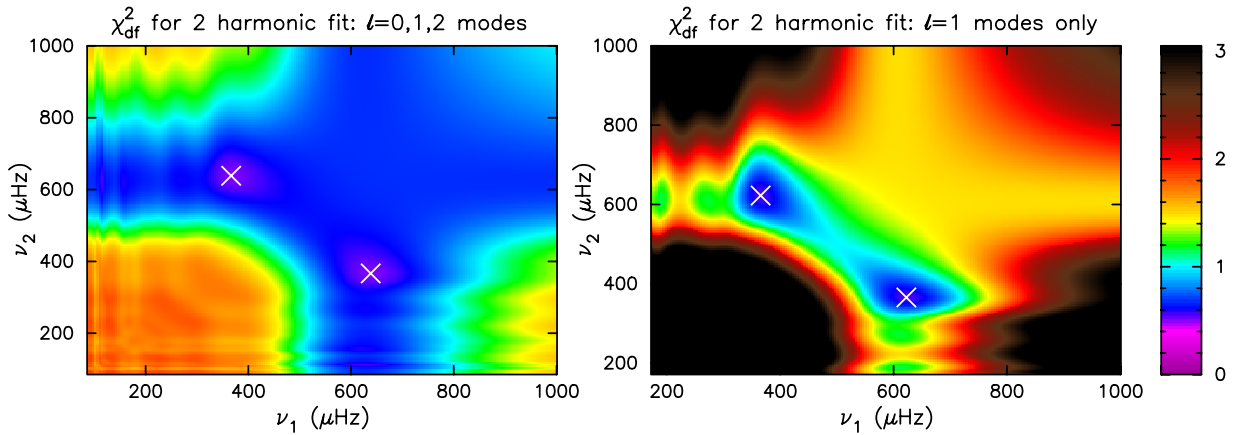, width=5.7cm,angle=270}
   \vskip-5pt 
\caption{HD49933:  $\chi_{df}^2$ for 2 harmonic fit to residuals of  $\beta_\ell$ to low order polynomial }
    \end{figure}  
\vskip 6pt
\parskip 2pt
\noindent{\bf References}

\footnotesize
\noindent Benomar O et al, 2009, A\&A, 507, L13\\
Roxburgh I W  \& Vorontsov S V, 2000, MNRAS, 317, 14:~~ 2002, ESA SP-485, p341, p337\\
Roxburgh I W  \& Vorontsov S V,  2003a, A\&A, 411, 215:  ~~~2003b, Ap\&SS, 284,187\\
Roxburgh I W, 2009, A\&A, 493, 185: ~~~~~~~~~~~~~~~~~~~~~~~~~~~ 2010, Ap\&SS, 328, 3\\
Vorontsov S V, 2001, ESA SP-464, 653

\end{document}